# The design of an agent based model of human activities and communications in cardiac resuscitation

Lyuba Mancheva
GIPSA-Lab / LPNC / Univ. of Grenoble Alps /
Laboratoire d'Etude des Mécanismes cognitifs,
Univ. of Lyon 2
Lyuba.Mancheva@gipsa-lab.grenoble-inp.fr

Julie Dugdale
Univ. of Grenoble Alps /
Grenoble Informatics Lab. (LIG) /
Agder Univ.
Julie.Dugdale@imag.fr

**Abstract**

*Cardio-pulmonary arrest is a common emergency situation causing over 400,000 deaths per year, more than a 1000 per day, in the USA alone. The goal of this work is to develop an agent based computer simulator that will allow trainers to experiment with different communication protocols, such as those found in air traffic control. This paper describes the first step in designing the simulator development. The design is based on an analysis of communications during real life training simulations using the FIPA standard categories.*

## 1. Introduction

Cardio-pulmonary arrest is defined as the sudden interruption of normal blood circulation due to the failure of the heart to contract effectively. The lack of oxygen to the brain gives rise to unconsciousness followed by abnormal or absent breathing. After a short time the victim develops irreversible brain lesions leading to permanent brain damage and eventually death [16]. This emergency situation calls for immediate treatment in the form of cardio-pulmonary resuscitation (CPR) that involves external cardiac massage and artificial respiration. Depending on the type of heart rhythm, an electric shock is given through the chest wall using a defibrillating machine in order to restore the normal heart contractions.

For the best chance of patient survival it is imperative that medical personnel are trained to react quickly they must carry out the correct procedures and function efficiently as a team. Typically rescue teams first learn the basic CPR procedure through studying manuals, etc. Participating in simulations and practicing on a manikin supplement this knowledge. In some medical training establishments a special manikin, that simulates breathing and that has a pulse and blood pressure, is connected to specialist software such as MicroSim Inhospital [20]. This system allows the trainers to finely tune the patient's symptoms and set up complicated scenarios [9]. Using manikins in quasi simulation situations allows the rescue team to practice their response procedures and to improve their performance facilitated by a debriefing session given by the trainers.

Whilst the traditional simulations using a manikin are very good for members of the rescue team to practice their response actions, such simulators are not as useful for trainers who wish to design and test the actual intervention protocols that the team will use; for example testing a new communication protocol. Since practicing with rescue teams is time-consuming and difficult to organize, the trainers would like a tool to assess and test new protocols before they are suggested as replacements to existing ones. One area that has been identified for improvement in CPR response is the communication between resuscitation team members. Intra team communication has been frequently identified as being sub-optimal and badly structured [11] [1]. Information may be misheard, lost, or may need to be repeated, wasting valuable seconds during a resuscitation. This coupled with other factors such as stress, tiredness, lack of individual experience and lack of experience in working together as a team may have potentially harmful consequences for the victim [27].

The overall goal of this work is to develop a computer based simulator that will allow trainers to experiment with different communication protocols and to assess the effect of factors such as tiredness, etc. on the performance of the resuscitation team. The work is focused on the in-hospital advanced life support (ALS) phase for adults. ALS is undertaken when the full resuscitation team arrives and it includes specific medical activities such tracheal intubation, cardiac monitoring, intravenous access and administration of medicines, etc. Basic Life Support (BLS) on the other hand is the phase of providing preliminary medical care until the main resuscitation team arrives to perform ALS.

This paper describes the first step in the design and development of the simulator, the approach taken and the rationale behind the work. The paper is structured as follows. Section 2 describes background work on intra group communication and other factors affecting CPR response, whilst section 3 explains the resuscitation algorithm followed by the teams. The design of the simulator and the approach adopted is described in section 4. Some first results are presented in section 5. Section 6 concludes this paper and discusses how the simulator will be developed in the future.

## 2. Background

Despite recent advances in training emergency medical personnel, the survival rate for victims of cardiac pulmonary arrest is very low, varying between 3 to 30 percent [17], [4]. Survival rate is influenced by factors such as early recognition of the problem, early CPR and effective defibrillation, etc. There are also less obvious factors that play an equally critical role in influencing patient survival. Factors relating to the individual rescuer such as stress, tiredness and the level of experience in dealing with a CPR emergency adversely affect cognitive functioning and reduce psychomotor responses [23], [27]. This increases the probability of error and also the time taken to perform the CPR. The efficiency of the CPR team is also affected by whether or not they have previously worked together, and the effectiveness of the leader [2]. Teams that are used to working together are more efficient (Joint Commission on the accreditation of healthcare organizations, cited in [28].

Several studies have shown that communication is a very important factor for successful medical team cooperation [1]. Communication is essential for building good team dynamics, ensuring cooperation, promoting leadership and preventing errors [19], [27], [13]. Lack of good communication can have fatal consequences; one study showed that 43% of medical errors are due to failed or inadequate communication [15]. Specifically within a CPR team Englebert and his colleagues found that communications were sub-optimal with only 56% of communications being audible and 44% being understandable [11]. Although ambient noise level can affect communications and the construction of situation awareness within a team [7], this was not considered to be the main problem. Instead, the study concluded that communications should be better structured; perhaps by employing an air traffic control (ATC) feedback system of aeronautics. Also  CPR training should address communication aspects as well as task oriented aspects. The idea of using ATC communication models was investigated, not in CPR, but in the problem of patient handover from surgery to intensive care [5]. These authors devised a handover protocol where team members had clearly identified tasks, where who speaks at each moment is well defined, where communications used a feedback confirmation mechanism, and where one person is responsible for ensuring situation awareness. One objective of our work is to assess the effect, using simulation, on the time taken to conduct CPR if such a protocol was put in place.

## 3. Resuscitation process

### 3.1. Basic procedure

Although there are minor differences in CPR practices between countries, resuscitation teams and first aiders generally follow procedures such as those laid down in the European Resuscitation Council Guidelines [12]. The basic algorithm for ALS is shown in figure 1.

**Figure 1: Advanced Life Support cardiac arrest algorithm [25]**

Typically when the resuscitation team arrives BLS is already underway. Whilst assessing what happened a member of the team takes over thoracic compressions (30 chest compressions followed by 2 ventilations – CPR 30:2). Since giving chest compressions is very tiring the team needs to coordinate alternating this activity. Artificial ventilation, using a laryngeal mask and bag, a bag mask or via tracheal intubation, is also started. The handover between possible mouth to mouth ventilation during BLS and instillation of the bag needs to be coordinated. The leader will also ask another team member to install an intravenous (IV) line for giving medicines. During this period the defibrillator is installed and its pads are attached to the victim in order to assess the cardiac rhythm. Some rhythms are unshockable, but if the rhythm is shockable the leader calls for an electric shock and for everyone to stand back from the victim. During this time compressions and ventilations are suspended. If there is no pulse after the shock, CPR is resumed. Every 2 minutes the rhythm is assessed and a shock is given. If there is no pulse after the third shock the leader asks for adrenaline to be injected. Another medicine, amiodarone, is given after the fourth shock if there is still no pulse. These injections continue, alternating between adrenaline and amiodarone, after each pulse-less shock. ALS is stopped if the patient shows signs of life or if the leader decides that further treatment is futile.

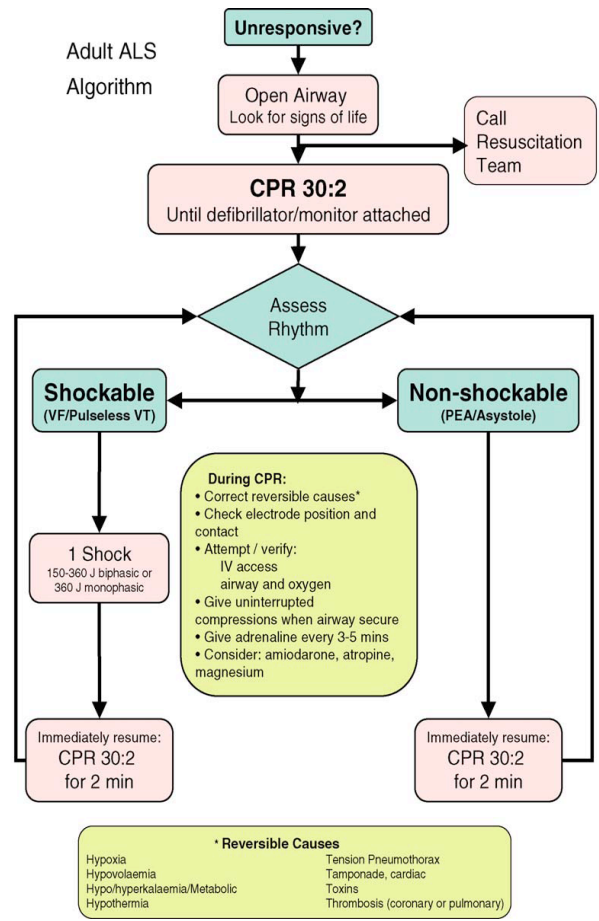

### 3.2. Analysis

Whilst the algorithm looks relatively sequential, in practice many activities occur in parallel and these need to be coordinated verbally by the designated leader of the team whilst he or she is receiving information on the patient's status from fellow team members. In essence the team leader is responsible for ensuring situation awareness and for directing the resuscitation procedure. Communication between team members may be directed to one specific member (e.g. the request to administer medicines to the person concerned with installing the IV line) or it may be directed to the whole group, e.g. issuing the instruction to stand back whilst the shock is given. Such broadcasted communications, either direct or indirect serve as way to increase situation awareness and increase team efficiency.

Referring back to the work of Catchpole [5] we can see similarities between how the CPR team works, and how the teams dealing with patient handover from surgery to intensive care work. In both cases team members have specific roles and one person is responsible for situation awareness. However, the communication protocol is currently very different. In the case of patient handover, who speaks at each moment is clearly defined and feedback mechanisms are employed to ensure that the message was received and understood. This type of protocol however has not been tried in CPR teams.

Fitzgerald Chase has proposed several changes to the communication protocol during resuscitation [14]. Firstly, in order to reduce the number of errors open communication via constructive intervention is proposed. This is where all the members of the team are encouraged to verbally challenge any decision of another team member, including the leader, if he or she feels that the current action is the wrong thing to do. At this stage in our work it was decided not to model this aspect. Secondly, closed-loop communication, where the intended listener repeats back the message for all to hear, should be used. This is similar to the approach used in ATC communication protocols. Finally, due to the time critical nature of resuscitation, team members have a tendency to all speak at the same time. Rather than aiding communication, this tends to be counter-productive, increasing the noise level and leading to confusion and misunderstandings. The suggestion is to rely more heavily on the team leader for coordinating and directing the communication. In summary, effective communication in emergency situations can be achieved through constructive intervention, closed loop communication, and defaulting to the team leader to manage the communications. Therefore, it is the effect of these changes to the communication protocol that the simulator aims to assess.

## 4. Approach and simulator design

Agent-based modeling (ABM) has seen an enormous growth in popularity in recent years. ABM is a subset of multi-agent systems that is concerned with developing a computational model for *simulating* the actions and interactions of a group of autonomous individuals. When the individuals represent people or groups then the term agent based social simulation (ABSS) is more often used. One of the main advantages of ABSS is that it gives us the opportunity to explore and experiment with social situations that we might not be able to do in real life. A simulation model can be set-up and then executed many times, varying the conditions in which it runs and exploring the effects of different parameters. These advantages, coupled with the difficulty in modeling social situations using a top-down mathematical approach, make ABSS a suitable choice for our particular problem.

The methodology adopted for this work follows an approach that has been used for many years for developing ABSS [6], [3], [21]. Two specific aspects separate this methodology from others used in ABSS; the first is in analyzing human behaviours and interactions through extensive field studies; the second is the strong emphasis given to validation of the construction of the model and the simulator. For the field studies, we attended; video and audio recorded several daylong practical training sessions and theoretical courses given at the Emergency Care Training Care Centre at the Grenoble Hospital, France. Our understanding was supplemented with extensive discussions and meetings with the trainers. In addition we also obtained some video data from the Mayo Clinic, Arizona, USA, of simulated real-life cardiac resuscitations that we analyzed to link the verbal communications to specific tasks.

The design of the simulator is the focus of this paper and it is described in the following sections by considering the agents and their interactions.

### 4.1. Agents

Each agent in the simulator represents a member of the resuscitation team. Currently we model the 4 basic team members: the physician and 3 paramedics or nurses, each of whom is responsible for different tasks. In the real situation there may be more members but the basic roles remain the same. Therefore, the limited number of agents in the current simulator does not greatly affect the design. We have limited the victim to be an adult as opposed to a child, pregnant woman or elderly person, since the resuscitation procedure is slightly different in these cases.

Agents have a number of actions or behaviours that they can do in response to the state of the patient or the environment. Certain actions are specific to the agent's role; for example Paramedic 2 performs all actions that concern respiration, such as intubating the patient. Other actions, such as checking the pulse or performing chest compressions, are generic and can be performed by all agents. Associated with each action is a time duration and probability of success in achieving the action. The time duration is implemented as a range and plausible ranges for each action were derived in consultation with the training physicians. It should be noted that medical staff with little experience would generally take longer to perform a task. Since the environment is non-deterministic, actions do not have a guaranteed effect. For example, there is no guarantee that the action of using the defibrillator on the patient will succeed in bringing about a normal heart rhythm. The probabilities of the success of actions were also derived with the training physicians. Note that the probability of some actions succeeding, for example doing chest compressions, is relatively high, reflecting the fact that performing chest compressions rarely fail to massage the heart. However other probabilities, such as that of a successful defibrillation may be quite low depending on other factors, such as the patient's age (age is modelled as an attribute of the agent), etc. The physical movements of agents around the patient are not modelled explicitly. From the videos and real-life simulations we found that the rescue team do not move around the patient very much.

The computational agents are reactive, responding to immediate changes in their environment such as a change in the patient's status. They are also proactive in the sense that they develop a plan to achieve their goals. This represents the idea that the agents proactively follow the ALS algorithm shown in figure 1. We have also used a Belief, Desire, Intention (BDI) approach [22]. Amongst other things, this allows the agent to have long-term goals (e.g. restore the patient's normal heart beat) and for each agent to have their own individual opinion or belief about the situation. Agents also have an auditory perception of their environment, which is implemented by the fact that agents can 'hear' verbal communications through message passing. Although non-verbal communication obviously plays a role in the real situation it is not currently implemented. An agent also has an attribute 'status' which is used to reflect if an agent is busy or available. If an agent is busy then it cannot, at that moment, undertake a task requested by another agent. The management of interruptions and resuming a task is similar to that in [7].

Characteristics such as level of stress, tiredness and experience are modelled as agent attributes, expressed in a range between 0 and 1. These levels, along with identifying which agent has worked with which other agent(s), are specified in the user interface and used as parameters in the simulation. The values for these characteristics probabilistically affect the time taken to do an action, thus they could reflect the idea that someone who is tired will probably take longer to do an action.

The patient is a particular kind of agent in that it cannot communicate with other agents; it only reacts to actions done upon it by other agents. Rather than model the detailed physiology of the patient, we model only the essential attributes: general health (a range from 0 to 1); breathing (Boolean); cardiac rhythm (range 0 to 150 for someone having had a cardiac arrest); and $CO_2$ (range 0, 100). Again these ranges were derived by consulting with the training physicians.

### 4.2. Interactions

During the simulation, agents send and receive messages to and from each other. The receipt of a message may provoke an agent to undertake an action, during the course of which the agent may send and receive other messages. The message structure contains the sender, the intended receiver, and the message content. In the case of broadcasted messages the receiver is a list of all other agents. We have used the standard ACL (Agent Communication language) of FIPA (Foundation for Intelligent Physical Agents). This has allowed us to classify communicative acts into the standard FIPA categories. So in addition to tracking the content of messages we can also see how many communications related to: error handling (errorHand), performing actions (PerfActions), requesting information (requestinfo), and passing on information (passingInfo), etc. Note that the text in parentheses refers to the FIPA predefined categories. As in real life there is the possibility that agents do not hear a message intended for it, or that the message was not well understood in which case the agent may ask the sender to repeat the message. As an example we could consider the communicative act of the leader (the physician) requesting the paramedic 1 to restart chest massage. This would be expressed as the following agent interaction: Request (phy, para1 chest_compressions (para1, patient), where Request is a FIPA communication category, phy (the physician) is the sender of the message, para1 (paramedic 1) is the receiver of the message, chest_compressions is the requested action that needs the parameters para1 and patient.

### 4.3. Experimental methodology

In order to further develop our simulator it is necessary to understand how people in a resuscitation team interact during a real CPR intervention. We could not collect data during a CPR with a human victim for ethical reasons. Therefore, we arranged with the local hospital training centre to collect data during 6 simulation exercises (6 different CPR teams). The centre regularly runs these exercises on a specially developed manikin in order refresh the training of their experienced staff. The aim of collecting the data was to see the different types of interactions (i.e. categorise real communications into the FIPA categories) and perform a quantitative and qualitative analysis of the exchanges. The results of this are reported in the next section. Once we understood how communications worked in the real world then we could try to replicate the situations in the simulator. Once we could ensure that the simulator was generating outputs similar to those generated in the real world, we could start to experiment with different communication protocols using the simulator. Such an approach is a standard way of calibrating and validating an agent based social simulator [8].

20 people, aged between 23 and 55, participated in the real-life simulations. 16 worked in specialist CPR teams, the others has been previously trained in CPR. Each team was composed of 4 people following the standard role configuration (i.e. 1 physician, 1 trained paramedic, 1 ambulance person, and 1 junior person). All of the simulations were audio and video recorded, and the physician, as the leader of the team, wore a personal microphone. To annotate the videos we used ANVIL[1]. The communications were categorised according to FIPA, and senders and receivers of the message were members of the team (physician, paramedic, etc.). In addition to the type of message (request, inform, etc.) we also examined the message content and classified it into one of 5 categories (medical action, time, medicine administration, status of the patient and status of the equipment). As examples performing chest compressions is classified as a medical action; giving adrenaline is medicine administration, and checking the pulse is checking the status of the patient. These categories have been used in a similar study [26] concerning communication.

### 4.4. Computer simulator

A first version of the simulator has been implemented using RePast S[2]. At the start of the simulation the user can specify the patient agent's parameters, and the parameters of the members of the rescue team (e.g. paramedic 1 is very tired, stressed and has no experience of working with the other team members). Once the simulation begins the user can see the communications between agents in the form of a directed graph. Communications trigger actions that the computational agents can do. All communications are logged in a file for later analysis. The actions or tasks that each agent can do are clearly specified as actions that are allowable by different types of agents. Concerning imposing an order in which the agents speak, in our design each communication act is triggered after or before an action dealing with the resuscitation. So for example to start inturbating the patient, the doctor will ask the nurse to set up the inturbation equipment. Upon receipt of this message the nurse will perform the action. By requiring the agents to send an acknowledgment message after receiving a communication we can test the efficiency of the feedback loop used in aviation ATC models.

---

[1] http://www.anvil-software.de/
[2] http://repast.sourceforge.net/
[3] No Flow is a critical measurement of the effectiveness of a team and refers to the time during which there is no cardiac debit, i.e. the time during which the patient is not receiving cardiac massage. The

Whilst these acknowledgement messages are likely to slightly increase the overall communication time, we hypothesize that the time gained from preventing conversations 'hanging' due to misunderstanding the message or not hearing it will be greatly reduced. Since the work is still in its early stages we have not yet tested this part.

Concerning time, each time step of the simulator represents 1 second in real time. This allows us eventually to compare the duration of resuscitation, and critically the duration of the 'No Flow'[3] time under different communication protocols.

## 5. First results

The physicians responsible for CPR training undertook a qualitative analysis of the performance of the real CPR teams during the real-life simulations. They classed their performance as being either good or bad, following 12 medical criteria and 50 communication criteria [24]. We also undertook a quantitative analysis. Figure 2 shows the number of communications by FIPA category, expressed as a percentage.

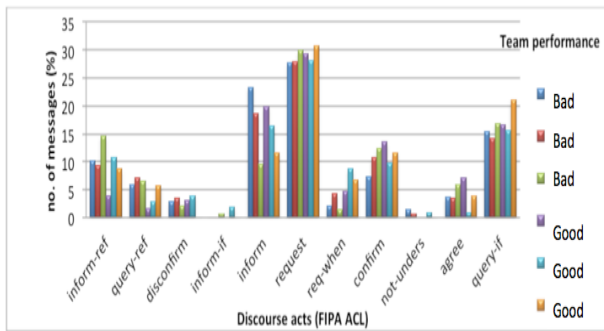

**Figure 2. Number of communications by FIPA category, expressed as a percentage**

From this figure we can see that the most common type of discourse for both good and bad teams was 'request' (i.e. asking someone to do an action). However what is interesting is that the teams that performed well, were more likely to use 'req-when' (i.e. asking someone to do an action when a condition is true) than teams that performed badly. This means that the good teams waited until a fact was known to be true before performing an action.

Upon further statistical analysis (T test and Mann-Witney) we found that the teams that performed well spoke more about medical actions, using more messages of the type 'perform an action'. This could indicate that they plan their actions together and work more as a team. Furthermore it confirms our hypothesis that to be efficient, tasks should be clearly assigned to specific members.

In addition the total number of messages exchanged was significantly less in teams that performed well as opposed to teams that performed badly. This confirms the findings of other studies that it is not necessary that everyone is aware of the situation, but rather that the right information arrives to the right person at the right time [18], [10]. In addition it may indicate that the team follows a well-defined order of speaking related to tasks.

Furthermore, we confirmed that well performing teams used more communications concerning 'req-when' and less 'query-ref' (ask a question concerning the value of a statement). Query-ref is usually used prior to an inform statement, i.e. an agent is questioning another agent on the value of something. This could indicate a lack of knowledge about the procedure to follow.

## 6. Conclusion and further work

We have presented the design of a simulator that in the future will be able to test different communication protocols. In addition we have presented the results of real-life simulations of CPR showing the type of communication according to the FIPA standard. The work is obviously in its early stages, however the basic implementation is in place to assess if a different communication protocol can reduce the CPR time and the no flow time. Since agents are proximally close they can overhear what other members of the team are saying. This has been implemented in the simulator by a can_hear function in the computational agent. However we are aware that communication is not always verbal and that non-verbal communications play an important role. As a first step in dealing with non-verbal communications we would like to implement a seeing behavior (function) in the agents, where if there is no obstruction an agent can observe the current actions of another agent. Since it has been found that a person performing an action may do it with the intention that other people see it, this forms a valid non-verbal communication mechanism.

## 7. Acknowledgements

---
[3] No Flow is a critical measurement of the effectiveness of a team and refers to the time during which there is no cardiac debit, i.e. the time during which the patient is not receiving cardiac massage. The value of No Flow should be close to zero.

This work has been undertaken as part of a collaboration between the Mayo Clinic in Arizona, USA, the Emergency Care Training Centre (CESU) which is part of the university hospital in Grenoble, France and the MAGMA research team in Grenoble. We are indebted Dr Bhavesh Patel from the Mayo clinic in Arizona who originally suggested investigating the problem of communication in resuscitation teams and who worked with us on the early versions of the model. We are also extremely thankful to Dr Koch from CESU who helped us greatly in understanding the medical aspects of cardiac resuscitation and who enthusiastically worked with us on the design. Julie Dugdale would also like to thank the University of Agder, to which she is affiliated.